\documentclass[aps, physrev, reprint, superscriptaddress]{revtex4-2}

\usepackage{graphicx}
\usepackage{xcolor}
\usepackage{comment}
\usepackage{dcolumn}
\usepackage[colorlinks=true,citecolor=blue,urlcolor=blue]{hyperref}
\usepackage[utf8]{inputenc}
\usepackage{xcolor}
\usepackage{amsmath}
\usepackage[T1]{fontenc}
\begin{document}

\title{Novel 2D Altermagnetic Vanadium Oxide with a Buckled Lieb Structure}

\author{Tamer Taşkıran}
\affiliation{UNAM - National Nanotechnology Research Center and Institute of Materials Science and Nanotechnology, Bilkent University, Ankara 06800, Turkey}

\author{Soheil Ershadrad}
\affiliation{Materials Design Division, Department of Physics, Chemistry, and Biology (IFM), Link\"oping University, SE-58183 Link\"oping, Sweden}

\author{Biplab Sanyal}
\affiliation{Department of Physics and Astronomy, Uppsala University, Box 516, 751 20 Uppsala, Sweden}

\author{Cüneyt Şahin}
\email[To whom correspondence should be addressed:\\]{cuneyt@unam.bilkent.edu.tr}
\affiliation{UNAM - National Nanotechnology Research Center and Institute of Materials Science and Nanotechnology, Bilkent University, Ankara 06800, Turkey}
\affiliation{Department of Physics and Astronomy, University of Iowa, Iowa City, Iowa 52242, USA}

\date{\today}

\begin{abstract}
Altermagnetism has recently emerged as a highly promising phase for spintronics, offering the combined advantages of both antiferromagnets and ferromagnets. Here, using a first-principles analysis based on density functional theory (DFT), we identify a monolayer V$_2$O crystal in a buckled Lieb lattice as a promising two-dimensional altermagnetic material. The structural and thermal stability of V$_2$O is verified through calculations of the crystal's formation energy, phonon structure, room-temperature ab initio molecular dynamics, and stiffness matrix. The system is found to exhibit auxetic behavior with a negative Poisson's ratio. Our calculations indicate an antiferromagnetic ground state with a local magnetic moment of $2.79\,\mu_{\mathrm{B}}$ per V atom and a magnetocrystalline anisotropy that favors an out-of-plane easy axis. The electronic structure exhibits a momentum-dependent spin splitting of 1.2 eV, which is a characteristic of altermagnets. Inclusion of spin-orbit coupling breaks the symmetry of the quadratic bad crossing near the Fermi level, resulting in a large Berry curvature and significant intrinsic spin Hall conductivity around $40\,(\hbar/e)\,\mathrm{S\,cm^{-1}}$. The results demonstrate that monolayer V$_2$O serves as a robust room-temperature altermagnetic platform, exhibiting magnetic anisotropy and spin-dependent transport responses. 
\end{abstract} 


\maketitle

\section{Introduction}

With the discovery of spin-polarized transport \cite{mott1936electrical}, spintronics has emerged as a prominent research field in which electron spin is used to store, process, and transmit information \cite{zutic2004spintronics}. In particular, it has driven magnetic storage technologies through phenomena such as giant magnetoresistance \cite{baibich1988giant,binasch1989enhanced}, magnetic tunnel junction \cite{julliere1975tunneling}, and spin-transfer torque \cite{slonczewski1996current, berger1996emission}. Spintronics has also become a viable solution for quantum computing, neuromorphic computing \cite{grollier2020neuromorphic}, and a low-power alternative to CMOS technology. So far, the main component of the spintonic devices has been ferromagnetic materials. However, they are limited by external stray fields and a relatively low frequency range. Antiferromagnets could be an alternative \cite{jungwirth2016antiferromagnetic}; however, they lack the emergent spin-polarized properties required for efficient readout and control. Recently, a third class of permanent magnets, namely altermagnets, has been discovered, fundamentally enhancing the traditional understanding of magnetic materials \cite{smejkal2022beyond, smejkal2022emerging}. Altermagnets as a third class of collinear magnetic phase are distinguished from these two classes of magnetic materials as the symmetry of the real space and the spin space are decoupled. Due to this decoupling, they can exhibit spin-split bands without requiring strong spin-orbit coupling, while maintaining compensated magnetic order in which time-reversal symmetry is broken, yet exhibiting zero net magnetization. As a result, altermagnets maintain the robustness of antiferromagnetic materials against external perturbations and stray fields, allowing them to generate highly spin-polarized currents with potential for terahertz-speed spintronic devices. They can also exhibit transport properties such as anomalous  Hall conductivity (AHC) and spin Hall conductivity (SHC), which are extremely important for electronic and spintronic applications. Therefore, they are promising for spintronic applications as they combine the advantages of both well-known ferro- and antiferromagnetic phases \cite{bai2024altermagnetism, jungwirth2025altermagnetic}. 

An important direction in the research landscape of altermagnetism is the discovery of novel two-dimensional materials. Although altermagnetism has been theoretically predicted and experimentally verified in three-dimensional crystals such as RuO$_2$ \cite{smejkal2020crystal}, CrSb \cite{Reimers2024}, and MnTe \cite{krempasky2024altermagnetic}, the discovery of stable two-dimensional altermagnets, which have the potential to be integrated into nanoscale devices, remains unrealized. The preliminary studies in this field are the high-throughput scanning of the Computational 2D Materials Database (C2DB) \cite{sodequist2024two}, the high-throughput scanning of the pentagonal 2D materials \cite{wang2026pentagonal}, engineering stacked bilayer systems \cite{pan2024general}, classifying 2D spin layer groups \cite{zeng2024description}, and the Janus type transition-metal oxychalcogenide monolayers \cite{ma2021multifunctional, zhu2024multipiezo, xie2025piezovalley, wu2024valley, guo2023piezoelectric} or artificial structures, CrO \cite{Chen2023}. Recently, the Lieb lattice has emerged as a possible structure that may host altermagnetic order. Originally, the Lieb lattice, which is a quasi-2D structure, was used as a toy model to examine altermagnetism in two-dimensional materials \cite{brekke2023two, durrnagel2025altermagnetic, kaushal2025altermagnetism}. Since then, it has been shown that the materials with Lieb lattice (or inverse Lieb lattice \cite{chang2026inverse}) structures could display topological properties \cite{antonenko2025mirror}, superconductivity \cite{brekke2023two, leraand2025phonon}, and piezo-response \cite{takahashi2025elasto, xu2025alterpiezoresponse, khodas2026tuning}. Once considered as a toy model, materials with Lieb lattices have been discovered, such as La$_2$O$_3$Mn$_2$Se$_2$ \cite{wei20252} and AV$_2$Se$_2$O \cite{jiang2025metallic, zhang2025crystal}.  Although most of these discovered materials included the transition-metal oxychalcogenide monolayers, there are also simpler transition-metal oxides with this structure, such as Cr$_2$O \cite{zhang2025giant, guo2026external} and Cr$_2$O$_2$ \cite{liu2026uncompensated}. Among those, transition metal oxides exhibit a wide variety of properties due to their different oxidation states and strongly correlated electrons, making them strong candidates for exploring emergent mechanical, electronic, and magnetic properties. A recent theoretical study on Cr$_2$O \cite{zhang2025giant} showed that the 2D Lieb lattice can be an ideal platform for a stable and robust intrinsic d-wave altermagnet.

In this paper, we propose and investigate another novel two-dimensional material using first-principle calculations, V$_2$O, that has a buckled Lieb lattice structure, which has not been discovered within the phase diagram of vanadium oxides (V$_x$O$_y$) \cite{szymanski2018electronic, hu2023vanadium}. First, we establish the dynamical stability of the V$_2$O monolayer by calculating the phonon dispersion and elastic properties, which reveal an anomalous negative Poisson ratio indicating an auxetic behavior. Next, we discuss its electronic and magnetic properties exhibiting altermagnetism. Our results show that a zero AHC (expected due to symmetry) and a robust spin Hall conductivity at the Fermi level are computed from the Berry curvature and the Kubo formula. We also discuss indicators of its topological properties, showing a Dirac-like crossing at the Fermi level.

\section{Computational Methods}
Density functional theory calculations were performed using the Vienna Ab initio Simulation Package (VASP) \cite{kresse1993ab, kresse1996efficient, kresse1996efficiency}. The interaction between the ion cores and valence electrons was described using the projector augmented-wave (PAW) method \cite{blochl1994projector}, and exchange–correlation effects were treated within the generalized gradient approximation (GGA) \cite{perdew1986accurate, perdew1986density} as parameterized by Perdew–Burke–Ernzerhof (PBE) \cite{perdew1996generalized}. Strong on-site Coulomb interactions for localized orbitals were accounted for using the DFT$+U$ approach within the Dudarev formalism \cite{dudarev1998electron}, where the effective Hubbard parameter $U=3.25$ eV was considered for the $d$-orbitals of V \cite{wang2006oxidation, moore2024high, horton2025accelerated}. A plane-wave energy cutoff of $10^{-8}$ eV was employed, and Brillouin zone integrations were carried out using a Gamma–centered Monkhorst–Pack k-point mesh of $24 \times 24 \times 1$. Vacuum length in the $z$-axis is set to 20 \AA. To further validate the electronic structure results, calculations using the screened hybrid functional HSE06 \cite{heyd2003hybrid} were performed and compared with those obtained from DFT+U. Due to the higher computational cost of the hybrid functional, a reduced Monkhorst–Pack k-point mesh of $8 \times 8 \times 1$ was employed for the HSE06 calculations. All other computational parameters were kept consistent to ensure a meaningful comparison between the two approaches. Formation energy calculations were also made using the hybrid approach. For the calculations of chemical potential of reference species an energy cutoff of 300 eV and a k-mesh of $8 \times 8 \times 8$ are used for body-centered cubic Vanadium crystal with lattice parameter 2.98 \AA, and energy cutoff 600 eV and a k-mesh of $1 \times 1 \times 1$ are used for O$_2$ molecule in a cubic box of length 8 \AA.

Structural relaxations were performed until the Hellmann-Feynman forces on each atom were less than $10^{-4}$ eV/\AA. Phonon dispersion calculations were carried out within the finite-displacement supercell approach using a $5 \times 5 \times 1$ supercell, as implemented in PHONOPY \cite{togo2023first, togo2023implementation}. With the same supercell, \emph{ab-initio} molecular dynamics (AIMD) simulations were conducted using VASP, the canonical (NVT) ensemble with a Nosé–Hoover thermostat \cite{nose1984unified, hoover1985canonical}, a 2.0 fs time step, and a constant temperature of 300 K. Elastic constants were calculated using the energy-strain method with strain increments of $5\times10^{-3}$. The Heisenberg magnetic exchange interactions ($J_{ij}$ and $D_{ij}$) were evaluated using the full-potential linear muffin-tin orbital (FP-LMTO) code RSPt \cite{wills2010full} within the Liechtenstein-Katsnelson-Antropov-Gubanov (LKAG) formalism \cite{liechtenstein1987local}. Monte Carlo simulations were performed using the UppASD code \cite{eriksson2017atomistic} to determine the magnetic ordering temperature $T_{N}$ based on the Heisenberg Hamiltonian. To obtain statistically converged properties, five independent ensembles were simulated using a $64\times64\times1$ supercell under periodic boundary conditions.

The Berry curvature and following anomalous and spin Hall conductivity calculations were carried out with the Wannier90 package \cite{pizzi2020wannier90} on top of the DFT$+U$ ground state calculations. Wannierization is done using the Marzari-Vanderbilt method \cite{marzari1997maximally}. Initial guess for the unitary matrix $A_{mn}^{\mathbf{k}}$ is done using the $d-$orbitals of V and $p-$orbitals of O. Disentanglement procedure is conducted on a window, the minimum eigenvalue up to $E_f+3.0$ eV, where a frozen window of $E_f-5.0$ to $E_f$ is used \cite{souza2001maximally}. A k-mesh of $300 \times 300 \times 25$ is used to calculate Berry curvature.

Pre- and post- processing of the calculation data is done using ASE \cite{hjorth_larsen2017atomic}, VASPKIT \cite{wang2021vaspkit},  and XCrySDen \cite{kokalj1999xcrysdennew, kokalj2003computer} packages .

\section{Results}
\subsection{Structural Properties}
\begin{figure}[htbp]
\begin{center}
    \includegraphics[width=\linewidth]{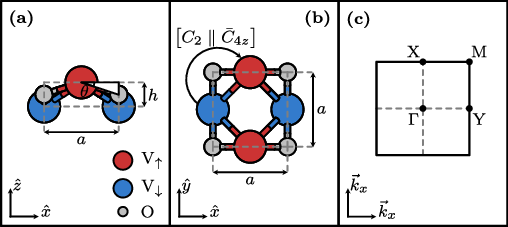}
    \caption{(a) The side view of the crystal structure of V$_2$O. The lattice parameters $a$, $h$, and $\theta$ are denoted. The V atom with the up spin is colored red, the V atom with the down spin is colored blue, and the O atom is colored gray. (b) The top view of the crystal structure of V$_2$O. The composite symmetry between two V atoms is denoted as $[C_{2} \parallel \bar{C}_{4z}]$, where $C_{2}$ is the $180^{\circ}$ rotation of the spin around an axis perpendicular to the spins and $\bar{C}_{4z}$ is the $180^{\circ}$ improper rotation symmetry between the lattice points of V atoms.  (c) The Brillouin zone and the high symmetry points of the V$_2$O.}
    \label{fig:crystal}
\end{center}
\end{figure}

We construct a V$_2$O monolayer with an inverse Lieb lattice structure by placing vanadium atoms on the edges of a square lattice and oxygen atoms at the corners. Structural optimization reveals that the lattice relaxes into a buckled configuration, in which one vanadium atom displaces upward along the $z$-direction while the other moves downward, as shown in Fig.~\ref{fig:crystal}. This buckling stabilizes the monolayer, analogous to silicene \cite{cahangirov2009two}. The relaxed structure has lattice parameter $a = 4.01$~\AA{} and buckling height $h = 0.54$~\AA{}, corresponding to a buckling angle of $7.7^\circ$. The resulting structure belongs to the tetragonal point group $\bar{4}m2$ ($D{2d}$) and space group $P\bar{4}m2$ (No. 115). The vanadium atoms occupy Wyckoff position $2g$, while the oxygen atoms occupy $1c$. The crystal symmetry includes a fourfold rotoinversion axis through the center of the square and mirror symmetries with respect to the (100) and (010) planes. When combined with a possible antiferromagnetic ordering in spin space, the system preserves a combined $[C_{2} \parallel \bar{C}_{4z}]$ symmetry, suggesting a $d$-wave altermagnetic state.

\begin{figure}[htbp]
\begin{center}
    \includegraphics[]{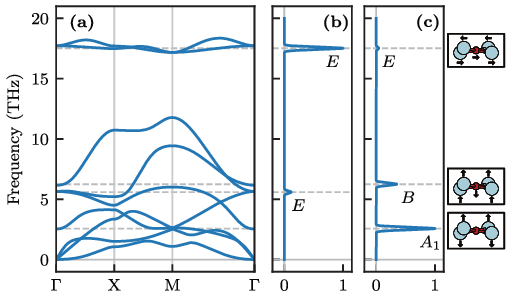}
    \caption{(a) Phonon dispersion of V$_2$O (b) IR spectrum of V$_2$O (c) Raman spectrum of V$_2$O and the vibration modes corresponding to the Raman peaks}
    \label{fig:phonon}
\end{center}
\end{figure}

The stability of this novel material is examined using four criteria. First, we analyze thermodynamic stability by calculating the crystal's formation enthalpy from the elements that comprise the compound. In this case, we use the following expression.
\begin{equation}
\label{eq:form}
    \Delta E_f = E_{\text{V$_2$O}} - \left[2 \cdot E_{\text{V-bulk}} + E_{\text{O-gas}}\right]
\end{equation}
Our calculation yields an enthalpy of formation, $\Delta E_f = -3.64$ eV, indicating that the crystal can form without disintegrating into its constituents.  The second criterion is the material's dynamic stability, assessed through the phonon dispersion relation. The phonon dispersion of V$_2$O calculated across the Brillouin zone exhibits no imaginary frequencies as shown in Fig. \ref{fig:phonon} (a), confirming the dynamical stability of the monolayer. The third criterion is the material's thermal stability. Through \emph{ab initio} molecular dynamics (AIMD) simulations, we demonstrate the thermal stability at room temperature; the atomic configuration remains preserved over a 1 ps duration with no structural reconstruction or bond breaking. The total energy exhibits only minor thermal fluctuations without systematic drift, supporting the feasibility of experimental realization under ambient conditions.

\begin{figure}[htbp]
\begin{center}
    \includegraphics[width=\linewidth]{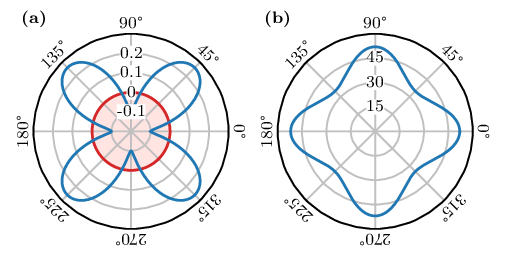}
    \caption{The angular dependence of (a) Poisson's ratio ($\nu_{\theta}$) and (b) Young's modulus ($Y_{\theta}$) as a function of the in-plane angle ($\theta$) for the V$_2$O monolayer. The auxetic region, where Poisson's ratio becomes negative, is highlighted by the red shaded area.}
    \label{fig:poisson}
\end{center}
\end{figure}

The fourth criterion we studied is the material's stiffness matrix. The calculated reduced stiffness matrix for the V$_{2}$O monolayer, which is constrained by the symmetry considerations of a two-dimensional square monolayer, is given by
\begin{equation*}
\mathbf{C}  =
\begin{bmatrix}
C_{11} & C_{12} & 0 \\
C_{12} & C_{22} & 0 \\
0 & 0 & C_{66}
\end{bmatrix} = 
\begin{bmatrix}
52.5 & -5.2 & 0.0 \\
-5.2 & 52.5 & 0.0 \\
0.0 & 0.0 & 13.8
\end{bmatrix} \mathrm{(N/m)} ,
\end{equation*}
Since all eigenvalues are positive, the stiffness matrix is positive definite. Therefore, the V$_{2}$O monolayer satisfies the mechanical stability criteria. Once we obtain the material's stiffness matrix, we can estimate several important mechanical properties. We start with the Poisson ratio, which is the transverse strain response of a material under uniaxial stress. The angle-dependent Poisson's ratio of the V$_2$O monolayer was evaluated using the following relation:
%
%
\begin{equation}
\nu(\theta)
=
\cfrac{
C_{12}(s^{4}_{\theta}+c^{4}_{\theta})
-
\left(2C_{11}-\cfrac{C_{11}^{2}-C_{12}^{2}}{C_{66}}\right)
s^{2}_{\theta}c^{2}_{\theta}
}{
C_{11}(s^{4}_{\theta}+c^{4}_{\theta})
-
\left(2C_{12} - \cfrac{C_{11}^{2}-C_{12}^{2}}{C_{66}}\right)
s^{2}_{\theta}c^{2}_{\theta}
}
\end{equation}
where $s_{\theta}$ denotes $\sin(\theta)$ and $c_{\theta}$ denotes $\cos(\theta)$. Figure~\ref{fig:poisson} (a) illustrates the angular dependence of Poisson's ratio ($\nu$) as a function of the in-plane angle ($\theta$) for the V$_2$O monolayer. The results reveal that the V$_2$O monolayer exhibits auxetic behavior, which means the material expands laterally when stretched and contracts laterally when compressed. Such materials are relatively rare but can exhibit unusual and attractive mechanical properties \cite{lakes1987response, greaves2011poissons, Chen2020}. The most negative value of the Poisson's ratio occurs along the $x$- and $y$-directions, where $\nu \approx -0.10$. In contrast, when the strain is applied along the diagonal direction ($\theta = 45^\circ$), Poisson's ratio becomes positive with a value of $\nu \approx 0.26$.

In a similar manner, the angle-dependent Young's modulus of the V$_2$O monolayer can be expressed as
%
%
\begin{equation}
Y(\theta)
=
\cfrac{C_{11}^{2}-C_{12}^{2}}
{C_{11}(s^{4}_{\theta}+c^{4}_{\theta})
-
\left(2C_{12}-\cfrac{C_{11}^{2}-C_{12}^{2}}{C_{66}}\right)
s^{2}_{\theta} c^{2}_{\theta} }.
\end{equation}
The angular dependence of Young's modulus is shown in Figure~\ref{fig:poisson} (b). The results reveal a pronounced anisotropic behavior. The maximum Young's modulus occurs along the $x$- and $y$-directions with a value of 52.0~N/m, while the minimum value of 34.9~N/m appears along the diagonal direction ($\theta = 45^\circ$).

The final analysis in this section is to determine the allowed IR and Raman peaks for experimental studies. The IR and Raman spectra are obtained from the DFT-computed vibrational modes at the $\Gamma$-point by evaluating their coupling to electromagnetic fields. The IR intensities shown in Fig. \ref{fig:phonon} (b) are determined from the Born effective charge tensors and the dielectric response, which quantify how each phonon mode induces a change in the macroscopic dipole moment; consequently, only modes with E symmetry are IR-active in accordance with the selection rules. In contrast, the Raman activities shown in Fig. \ref{fig:phonon} (c) are derived from the change in the polarizability (dielectric tensor) with respect to atomic displacements along each normal mode. This leads to Raman-active contributions from $A_1$, $B$, and $E$ modes, each corresponding to distinct symmetry-preserving deformation patterns, as illustrated in the figure. The normalized intensities reflect the relative strength of light–matter interaction for each mode, and the complementary nature of IR and Raman activity is consistent with the crystal symmetry, where different irreducible representations govern dipole versus polarizability responses.

\subsection{Magnetic Properties}

\begin{figure}[htbp]
\begin{center}
    \includegraphics[width=\linewidth]{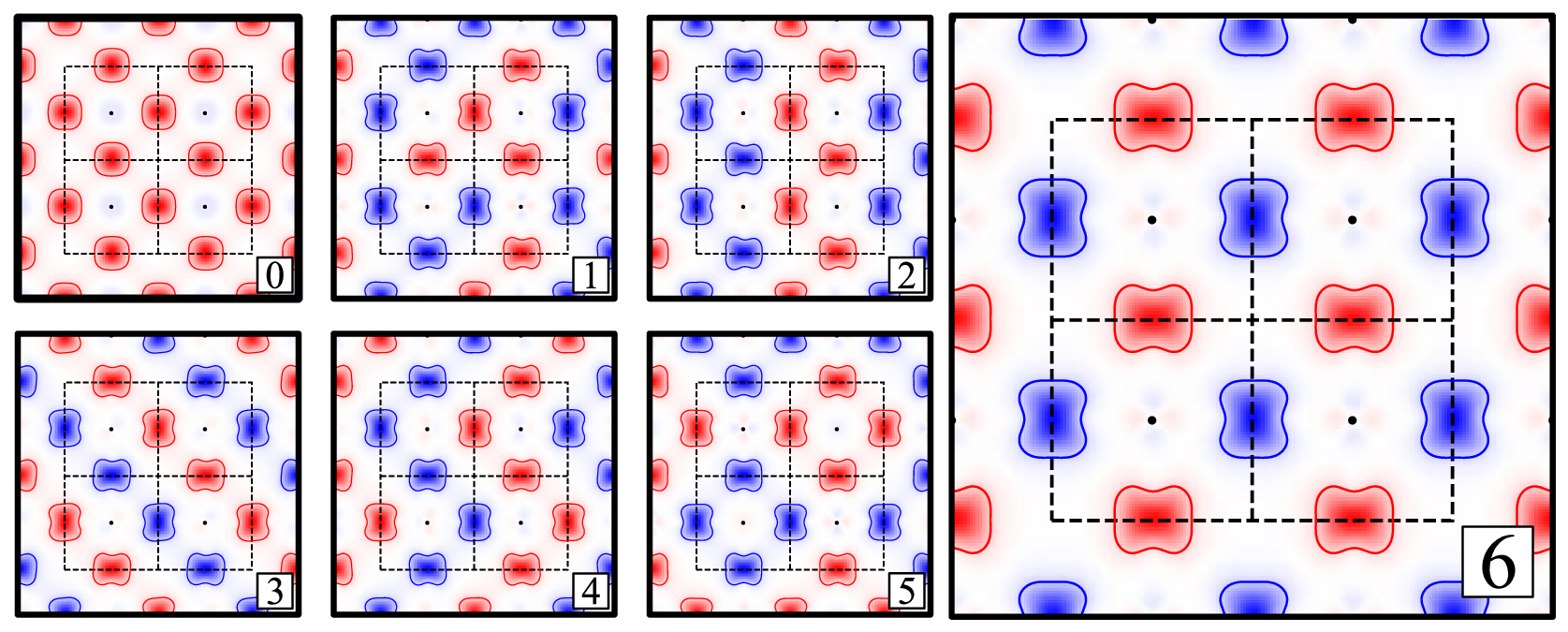}
    \caption{Spin density plot of spin-polarized DFT$+U$ calculations of different AFM configurations. Iso-contours are drawn at $\pm 0.00025$ $e/$\AA . Oxygen atom locations are indicated by black circles for visual clarity. The $2\times2\times1$ super-cell is shown with the dashed lines.}
    \label{fig:config}
\end{center}
\end{figure}

In this section, we analyzed the material's magnetic properties. The first step is to determine the magnetic ground state. We considered 6 different configurations, each representing a different magnetic space group among all possible antiferromagnetic configurations of $2\times2\times1$ supercell and the ferromagnetic configuration. To find the energetically favorable configuration, we calculated the energies of one configuration from each MSG within the spin-polarized DFT$+U$ formalism. The calculated spin densities of these configurations are shown in Fig. \ref{fig:config}. The results indicate that the striped antiferromagnetic configuration has the lowest energy state. In this state, there exists a local magnetic moment of $2.79~\mu_{B}$ per V atom with a charge of 4.15, while O atoms have no magnetic moment and have a charge of 5.06. Benchmarking against the hybrid HSE06 functional showed good agreement, with only 0.36\% differences in the atomic charges of V and O and 3.75\% in the magnetic moment of V, validating the $U=3.25$ eV parameter. 

Next, we aim to investigate the material's magnetic anisotropy. We calculate the energy state of the material with respect to the N\'eel vector, whose direction can be defined by two spherical angles, the angle between the N\'eel vector and its projection to the $xy-$plane (denoted by $\alpha$), and the angle between the N\'eel vector and the $z-$axis (denoted by $\beta$). We include the spin-orbit interaction in DFT$+U$ calculation to observe the effect of the alignment of the N\'eel vector. We repeat this calculation for 7 points in $\beta \in \left[0,\ \pi\right]$ and 6 points in $\alpha \in \left[0,\ \pi\right)$ and we fit the calculated data to the following equation of magnetic anisotropy energy (MAE):
\begin{equation}
\label{eq:MAE}
    E_{\text{MAE}} \approx K_1 \sin^2 \beta + K_2 \sin^4 \beta + K_3 \sin^4 \beta \cos 4\alpha
\end{equation}
Obtained values for $K_1 = 0.15 $ meV, whereas $K_2$ and $K_3$ are nearly zero. These numbers show that the system has the lowest energy state when the Néel vector is aligned along the $z-$axis. We also observe that the system behaves isotropically in the $xy-$plane when the Néel vector is not aligned with the $z-$axis. This clearly indicates that this crystal has an out-of-plane easy axis. The magnitude of the MAE is moderate compared to similar materials \cite{webster2018strain, fuh2016newtype}, indicating the tunability of the easy axis while maintaining stability.

\begin{figure}[htbp]
\begin{center}
    \includegraphics[]{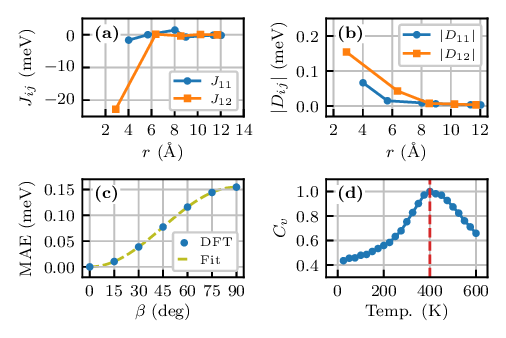}
    \caption{(a) Symmetric exchange interaction parameters with respect to distance (b) Asymmetric exchange interaction (DMI) parameters with respect to distance  (c) Magnetic anisotropy energy of V$_2$O. Blue markers show the results of spin-orbit interaction, including DFT$+U$ calculations for $\alpha=0$. Green curve shows the results of Eq. \ref{eq:MAE} for $\alpha=0$ (d) Heat capacity with respect to temperature, calculated through Monte Carlo simulation of the Heisenberg Hamiltonian.}
    \label{fig:exchange}
\end{center}
\end{figure}
We further evaluated the isotropic symmetric exchange interaction ($J_{ij}$) and the Dzyaloshinskii-Moriya interaction (DMI) vector ($\vec{D}_{ij}$) between the $i$-th and $j$-th lattice sites. Figures \ref{fig:exchange}(a) and (b) present the evolution of $J_{ij}$ and the magnitude of $\vec{D}_{ij}$ as functions of the pair distance, respectively. The antiferromagnetic (AFM) ordering is found to originate primarily from the nearest-neighbor inter-sublattice interaction ($J_{12}$), which exhibits a strong AFM coupling of approximately $-23$ meV. As the pair distance increases, the magnitude of $J_{12}$ decays rapidly, consistent with the behavior expected for gapped magnetic systems. In contrast, the intra-sublattice interaction ($J_{11}$) remains weak over the entire distance range and is therefore expected to contribute only marginally compared with $J_{12}$.
The non-centrosymmetric crystal structure of V$_2$O gives rise to finite, non-canceling DMI interactions, as illustrated in Fig.~\ref{fig:exchange}(b). The first-nearest-neighbor interaction $|D_{12}|$ is larger than $|D_{11}|$; however, the overall magnitude of the DMI remains relatively small (below $\sim 0.2$ meV) due to the weak spin-orbit coupling (SOC) strength of the V and O atoms. Given the small $D/J$ ratio of less than 0.01, the formation of skyrmions in V$_2$O is unlikely, particularly in the absence of an external magnetic field.
By incorporating the calculated exchange interactions ($J_{ij}$), Dzyaloshinskii-Moriya interaction vectors ($\vec{D}_{ij}$), and single-ion magnetic anisotropy constants ($K_i$) into the following spin Hamiltonian,
\begin{equation}
H = -\sum_{i\neq j} J_{ij}\vec{S}_{i}\cdot\vec{S}_{j}
-\sum_{i\neq j} \vec{D}_{ij}\cdot(\vec{S}_{i}\times \vec{S}_{j})
- \sum_{i} K_{i}(S_{i}^{z})^{2},
\label{H}
\end{equation}
where $\vec{S}_i$ and $\vec{S}_j$ denote the spin vectors at lattice sites $i$ and $j$, respectively, we performed Monte Carlo simulations to estimate the Néel transition temperature of V$_2$O.
Figure~\ref{fig:exchange}(d) presents the normalized heat capacity ($C_v$) as a function of temperature. A broad peak, characteristic of antiferromagnetic systems, is clearly observed, with its maximum corresponding to a magnetic transition temperature of approximately 400 K. This relatively high Néel temperature suggests that V$_2$O is a potential candidate for an above-room-temperature two-dimensional magnetic material. Elevated magnetic transition temperatures exceeding 300 K have also been experimentally reported in other V-based two-dimensional magnetic materials, such as VSe$_2$~\cite{bonilla2018strong}.

\begin{table}[t]
\caption{\label{tab:config}%
Magnetic Space Groups (MSG) and relative total energies ($\Delta E$) for various magnetic configurations, which are shown in Fig. \ref{fig:config}. The energy differences are reported in eV relative to the ground state (State 6, Striped AFM).}
\begin{ruledtabular}
\begin{tabular}{lllc}
\textrm{Config.}&
\textrm{Description}&
\textrm{MSG}&
\multicolumn{1}{c}{\textrm{$\Delta E$ (eV)}}\\
\colrule
0  & Ferromagnetic          & 115.283   & 3.97 \\
1  & AFM (Trigonal)         & 115.286   & 0.37 \\
2  & AFM (Double Striped)   & 115.289   & 0.60 \\
3  & AFM (Diagonal Striped) & 17.13     & 0.23 \\
4  & AFM (Zigzag)           & 28.93     & 0.43 \\
5  & AFM (Diamond)          & 5.15      & 0.63 \\
6  & AFM (Striped - N\'eel)          & 6.21      & 0.00 \\
\end{tabular}
\end{ruledtabular}
\end{table}

\subsection{Electronic Properties}

\begin{figure}[tbp]
\begin{center}
    \includegraphics[width=\linewidth]{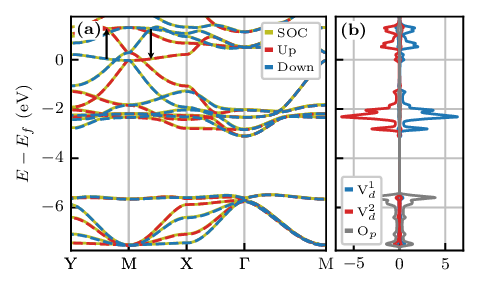}
    \caption{(a) Band structures of spin-polarized collinear DFT calculations vs SOC included noncollinear DFT calculations. Up and down bands of the spin-polarized results are plotted with blue and red lines, while SOC-included bands are drawn with dashed green lines. The couple of up and down bands with the largest spin-splitting is shown with the black arrows, whose direction mirrors around the $\mathrm{M}$ point. (b) Projected density of state results for the spin-polarized calculation. For simplicity, only the $d-$orbitals of the individual V atoms and the $p-$orbital of the O atom are shown.}
    \label{fig:v2o_band}
\end{center}
\end{figure}

In this part, we report the electronic properties of altermagnetic V$_2$O. Firstly, we calculated the spin-polarized band structure of the material with and without the spin-orbit interaction. We observe a clear momentum-dependent spin splitting of the bands along $\mathrm{Y}-\mathrm{M}-\mathrm{X}-\Gamma$ path where the bands along $\Gamma-\mathrm{M}$ path remain degenerate (i.e., it is the altermagnetic nodal line), as shown in Fig. \ref{fig:v2o_band} (a). Considering that this splitting is observed even without the SOC shows that the source of this splitting is the altermagnetic symmetry of the crystal, especially $[C_2 \parallel S_{4z}]$. Next, we quantify the splitting value as the energy difference between the corresponding up and down bands, i.e., $E_{\uparrow}\left(\mathbf{k}\right) - E_{\downarrow}\left(\mathbf{k}\right)$. We observe an energy splitting of approximately $1.2$ eV, which is notably large compared with other well-known altermagnetic materials \cite{smejkal2022emerging}. To further investigate the electronic structure, we also computed the projected density of states. The results for the $d-$orbitals of the individual V atoms and the $p-$orbitals of the O atom are shown in Fig. \ref{fig:v2o_band} (b). We observe that the electronic states near the Fermi level are predominantly derived from $d-$orbitals of the V atoms, with only a minor contribution from O$-p$ orbitals, while in the energy range below the Fermi level (approximately -6 to -2 eV), a noticeable overlap between V$-d$ and O$-p$ states is observed, indicating moderate hybridization between V and O atoms. Furthermore,  the difference between spin up and spin down density of states of the individual $d-$orbitals of V atom shows the strong localization of the $d-$orbitals of V atom and its magnetic moment.

\begin{figure}[h]
\begin{center}
    \includegraphics[width=\linewidth]{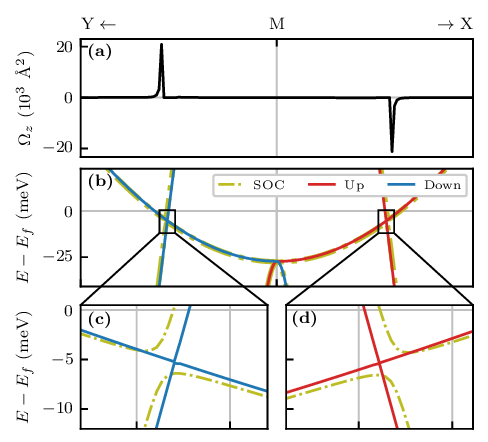}
    \caption{(a) Berry curvature plotted along the path of the high symmetry points $\mathrm{Y}-\mathrm{M}-\mathrm{X}$. The peaks coincide with the locations of the SOC-gapped Dirac crossings. (b) Band structure plotted along the same path, showing SOC included and spin-polarized band structure calculations. (c-d) Zoom in on the Dirac crossings, emphasizing the gapping caused by the SOC}
    \label{fig:v2o_soc}
\end{center}
\end{figure}

\begin{figure}[htbp]
\begin{center}
    \includegraphics[width=3.0in]{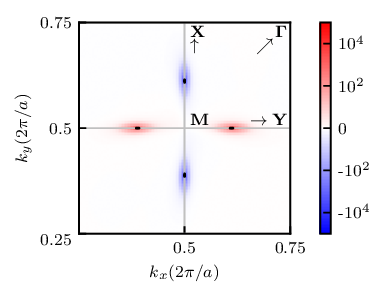}
    \caption{Berry curvature ($\Omega_z$) of V$_2$O in the units of $\text{\AA}^2$, plotted over the $k_z=0$ plane. The color map denotes the value of the Berry curvature. Black circles that coincide with the Berry curvature hot spots denote the Fermi surface at $k_z=0$. Directions of the high symmetry points are annotated on the figure. Both the Berry curvature hot-spots and the Fermi surface are strongly correlated to the altermagnetic symmetry of the crystal.}
    \label{fig:v2o_berry}
\end{center}
\end{figure}

Another important result the band structure shows is the existence of Dirac points with quadratic band crossings at the Fermi level along the $\mathrm{X}-\mathrm{M}$ and $\mathrm{M}-\mathrm{Y}$ paths, which are protected by mirror symmetry. This behavior is consistent with expectations for altermagnetic order on a two-sublattice square lattice, where Dirac-like crossings can emerge along mirror-symmetric lines in the absence of spin–orbit coupling (SOC) \cite{antonenko2025mirror}. Upon inclusion of SOC, the overall band structure remains largely unchanged; however, a gap opens at these crossing points. This gap opening is confirmed by a detailed inspection of the band structure in the vicinity of the Dirac points, as shown in Fig. \ref{fig:v2o_soc} (b-d). Introducing a gap at the Dirac points associated with a quadratic band crossing is known to yield Chern bands \cite{sun2009topological}. Therefore, we expect strong Berry curvature hotspots around these Dirac points. To verify the presence of large hotspots, we calculate the Berry curvature using MLWFs within the Kubo-Green formalism as follows.
\begin{equation}
    \Omega_{xy}^z(\mathbf{k}, \omega) = -\hbar^2 \sum_{n, m} \frac{2 \operatorname{Im} \left[ \langle \psi_{n\mathbf{k}} | A_x^z | \psi_{m\mathbf{k}} \rangle \langle \psi_{m\mathbf{k}} | v_y | \psi_{n\mathbf{k}} \rangle \right]}{(\epsilon_{n\mathbf{k}} - \epsilon_{m\mathbf{k}})^2 - (\hbar\omega + i\eta)^2}
\end{equation}
where $ A_x^z=\frac{p_x}{m}$  and $ A_x^z=\frac{\left[p_x, s_z\right]}{m}$ are velocity operators for the anomalous and spin Hall conductivities, respectively.. The Berry curvature calculated along $\mathrm{Y}-\mathrm{M}-\mathrm{X}$ path (shown in Fig. \ref{fig:v2o_soc} (a)) shows that two anti-symmetric peaks occur at the location of the Dirac points as expected. The Berry curvature calculated over the Brillouin zone plane $k_z=0$  (shown in Fig. \ref{fig:v2o_berry}) has the quadrupole structure that is consistent with the crystal symmetry, such that, 
\begin{equation}
    \Omega \left(k_x, k_y \right) = -\Omega \left(-k_y, k_x \right)
\end{equation}

\begin{figure}[htbp]
\begin{center}
    \includegraphics[width=\linewidth]{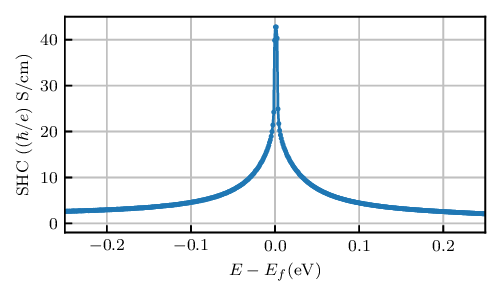}
    \caption{Spin Hall conductivity of V$_2$O with respect to the Fermi level.}
    \label{fig:v2o_shc}
\end{center}
\end{figure}

Considering the existence of Berry curvature hot spots in the reciprocal space, we might expect to observe Hall properties such as Anomalous Hall Effect (AHE) and Spin Hall Effect (SHE) \cite{Sahin2015}. Within the Kubo-Greenwood formalism, we can calculate the corresponding conductivities using the Berry curvature as follows.
\begin{equation}
    \sigma_{xy}^{z} = -\frac{e^2}{\hbar} \frac{1}{N_k V_c} \sum_{n, \mathbf{k}} f_{n \mathbf{k}} \Omega^z_{n, xy}(\mathbf{k}).
\end{equation}
Our AHC calculations result in zero conductivity, consistent with the constraints imposed by the magnetic point group symmetry and the out-of-plane Néel vector, according to the Landau theory of altermagnetism \cite{schiff2025collinear, mcclarty2024landau}.  On the other hand, the material exhibits a robust and non-vanishing SHC. Our transport analysis reveals that the SHC peaks prominently at the Fermi level, reaching a magnitude suitable for detection in a 2D material \cite{Sahin2019}, before rapidly decaying as the chemical potential shifts into the lower valence or higher conduction bands, as shown in Fig. \ref{fig:v2o_shc}. A zero AHC and a significant SHC indicates possibility of generating pure spin currents without causing a charge Hall voltage. In the clean limit, V$_2$O exhibits a remarkable spin current generation capacity without requiring doping or electrostatic gating that is revealed by the topological feature originating intrinsically from the electronic band structure.

\section{Conclusions}
In summary, we have proposed a novel two-dimensional material, V$_{2}$O, that crystallizes in a buckled inverse Lieb lattice. Our first-principles calculations demonstrate that this material is thermodynamically, dynamically, and mechanically stable. Notably, this monolayer exhibits an anomalous negative Poisson’s ratio of $\nu \approx -0.10$ along the $x$ and $y$ directions, identifying it as a rare 2D auxetic material. Magnetically, it hosts a robust d-wave altermagnetic state characterized by a striped antiferromagnetic ground state and an out-of-plane easy axis. The calculated Néel temperature of approximately 400 K indicates that the magnetic order remains stable well above room temperature, a critical requirement for practical applications. The electronic structure reveals a massive momentum-dependent spin splitting of up to 1.2 eV, driven by the crystal symmetry rather than spin-orbit coupling. Furthermore, our transport analysis indicates that the inclusion of SOC gaps the Dirac-like crossings at the Fermi level, leading to significant Berry curvature hotspots. While the anomalous Hall conductivity remains zero as dictated by symmetry, the material displays a robust and non-vanishing spin Hall conductivity that peaks at the Fermi level. These combined mechanical and topological properties make V$_{2}$O a highly promising platform for the next generation of high-speed, nanoscale altermagnetic and spintronic devices.  

\begin{acknowledgments}
This work was supported by the Scientific and Technological Research Council of Turkey (TÜBİTAK) under the 1001--Scientific and Technological Research Projects Funding Program (Project No. 124F108). The numerical calculations were carried out on the MareNostrum 5 supercomputer of Barcelona Supercomputing Center (BSC) and the TUBITAK ULAKBIM High Performance and Grid Computing Center (TRUBA). This work was supported by the BAGEP Award of the Science Academy.
B.S. acknowledges financial support from Swedish Research Council (grant no. 2022-04309) and STINT Mobility Grant for Internationalization (grant no. MG2022-9386). The computational resources provided by the National Academic Infrastructure for Supercomputing in Sweden (NAISS) at NSC(NAISS 2025/3-68) and PDC (NAISS 2025/3-68 and NAISS 2026/1-46) partially funded by the Swedish Research Council through grant agreement no. 2022-06725 are gratefully acknowledged.
\end{acknowledgments}

\bibliography{1-V2OPAPER}

\end{document}